\documentstyle[11pt,newpasp,twoside,epsf]{article} 
\def\edcomment#1{\iffalse\marginpar{\raggedright\sl#1\/}\else\relax\fi} 
\reversemarginpar 
 
\begin{document} 
\title{Integral-field Spectroscopy of Galactic Nuclei} 
\author{Tim de Zeeuw} 
\affil{Leiden Observatory, The Netherlands} 
 
\begin{abstract}
High-resolution imaging and long-slit spectroscopy obtained with HST,
combined with ground-based integral-field spectroscopy, provides the
kinematics of stars and gas in nearby galactic nuclei with sufficient
accuracy to derive the intrinsic dynamical structure, and to measure
the mass of the central black hole. This has revealed that many nuclei
contain decoupled kinematic components and asymmetric structures, and
that nuclear and global properties of galaxies are correlated.  Higher
spatial resolution and significantly increased sensitivity are
required to cover the full range of galaxy properties and types,
including the nearest powerful active radio galaxies, and to study the
evolution of galactic nuclei as a function of redshift.  The prospects
in this area are discussed.
\end{abstract}

\section{Integral-field spectroscopy}

The past decade has seen a revolution in instrumentation for spatially
resolved spectroscopy of galaxies and their nuclei. At many
observatories, traditional long-slit spectrographs have been replaced
by integral-field devices which produce spectra over an area, fully
spatially sampled, in many cases taking advantage of adaptive optics
capabilities (see, e.g., Emsellem \& Bland--Hawthorn 2002 for a recent
summary).  These instruments are very efficient in use of telescope
time, and allow complete reconstruction of the intensity distribution
from the spectra, so that errors in the registration of the spectra
relative to the galaxy image, familiar from aperture and long-slit
spectroscopy, are avoided.

\section{Galactic nuclei with HST} 

HST has revealed that most early-type galaxies have centrally cusped
luminosity distributions, often containing nuclear stellar and gaseous
disks as well as asymmetries. {\tt FOS} aperture spectroscopy and {\tt
STIS} long-slit spectroscopy of stars and gas has demonstrated that
nearly all these nuclei contain a supermassive black hole, with masses
ranging between $10^6$ and a few times $10^9 M_\odot$ (e.g., Kormendy
\& Gebhardt 2002).  Global and nuclear properties appear to
correlate. Examples include a correlation of cusp-slope with total
luminosity, and of black hole mass with host galaxy velocity
dispersion (e.g., Gebhardt et al., 2000, Ferrarese \& Merritt 2000).
We discuss four examples of HST studies of nuclei to illustrate the
need for two-dimensional spectroscopic coverage, the need for
measurement of the absorption-line kinematics, and the need for high
spatial resolution.

\begin{figure}
\vskip -8.0truept
\plotfiddle{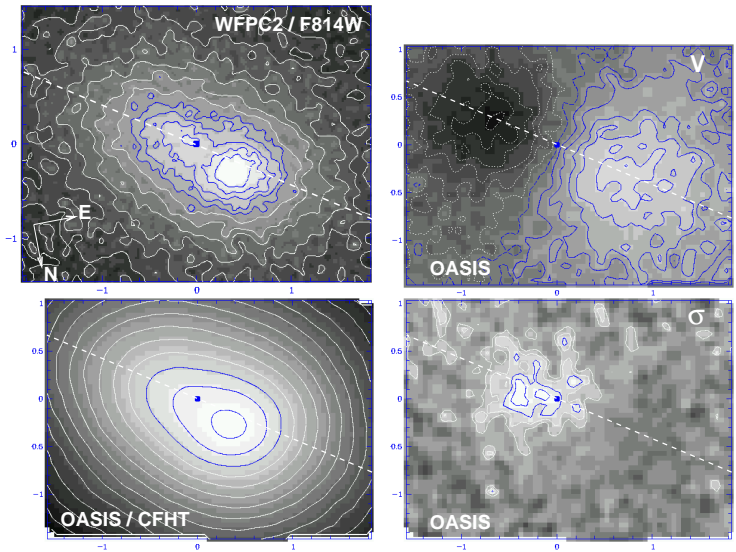}{7.5truecm}{0.0}{30}{30}{-145.0}{0.0}
\caption{The nucleus of M31. Clockwise from top left: deconvolved {\tt
         WFPC2} $I$-band image, stellar velocity $V$, stellar velocity
         dispersion $\sigma$, and intensity $I$ derived from {\tt
         OASIS} integral-field spectroscopy near the Ca triplet (Bacon
         et al.\ 2001a). The brightest peak in the {\tt WFPC2} image
         is P1, the secondary peak is P2. The dashed line indicates
         the symmetry axis of the $V$-field. The field-of-view is
         3\farcs5 by 2\farcs5. }
\vskip -8.0truept
\end{figure}

\subsection{The nucleus of M31} 
 
Stratoscope II discovered that the nucleus of M31 is asymmetric
(Light, Danielson \& Schwarzschild 1974). HST imaging revealed two
peaks in the brightness distribution, labeled P1 and P2 by Lauer et
al.\ (1993), separated by 0\farcs49.  The ground-based integral-field
spectroscopy with modest spatial resolution obtained with {\tt TIGER}
showed a nearly symmetric stellar velocity field, with P1 offset from
the kinematic center, and a velocity dispersion peak near P2 but
offset from it in the direction away from P1 (Bacon et al.\
1994). Multiple {\tt FOS} pointings (Ford, unpubl.), and long-slit
spectroscopy with both {\tt FOC} (Statler 1999) and {\tt STIS} (GTO
8018, PI Green, see Bacon et al.\ 2001a) showed an asymmetric velocity
curve, and confirmed the velocity dispersion peak near P2, presumably
caused by a $7\times 10^7 M_\odot$ black hole.  AO-assisted
integral-field spectroscopy with {\tt OASIS} on the CFHT, with an
effective resolution of 0\farcs45, clarified these puzzling
asymmetries (Bacon et al.\ 2001a).  As Figure 1 shows, the velocity
field is regular, and the dispersion peaks near P2, but a slit through
P1 and P2 (a natural choice based on the {\tt WFPC2} image) neither
coincides with the kinematic symmetry axis nor intercepts the true
$\sigma$ maximum: without the {\tt OASIS} data, the {\tt STIS}
profiles are difficult to understand.  The nature of the M31 nucleus
is not fully understood, but the N-body models of Bacon et al.\
(2001a), which build on Tremaine's (1995) eccentric disk model, show
that a near-Keplerian $m=1$ density wave provides a good qualitative
fit. Future models will have to deal with all the constraints
established from two-dimensional spectroscopy.

\begin{figure}
\plotfiddle{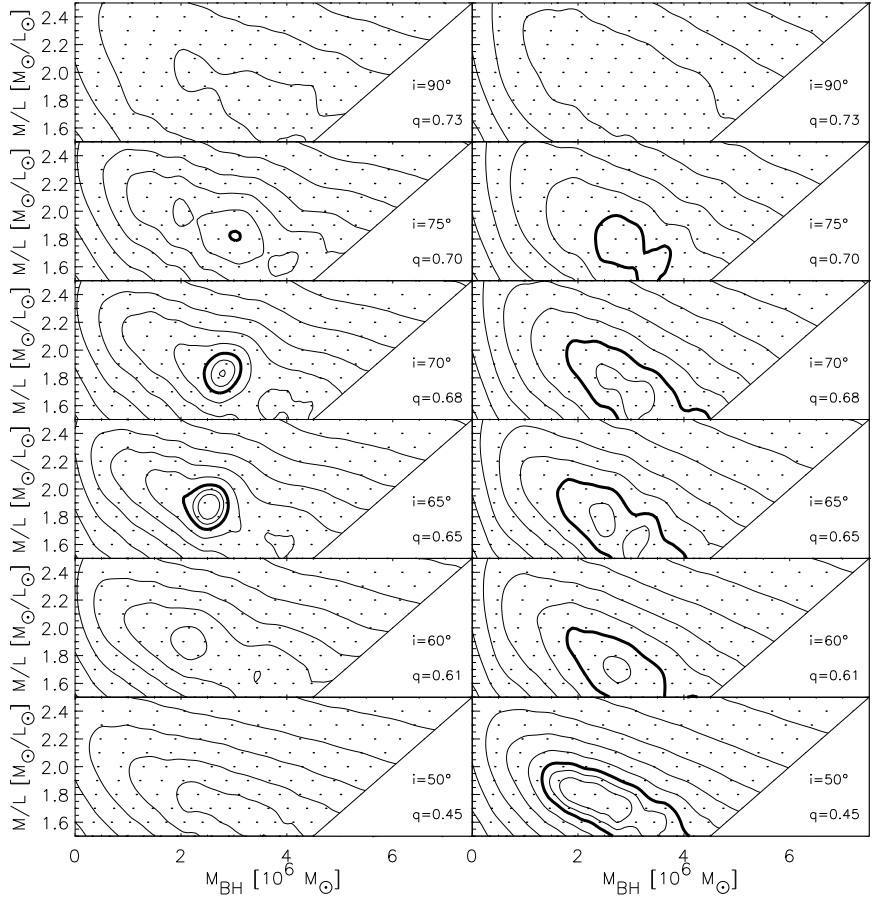}{12.0truecm}{0.0}{100}{100}{-180.0}{-30.0}
\caption{Dynamical models for M32. The panels show contours of
$\Delta\chi^2$, which measures the goodness of fit of the dynamical
models to the kinematic measurements, as a function of black hole mass
$M_{\rm BH}$, stellar mass-to-light ratio $M/L$ and inclination $i$,
taken from Verolme et al.\ (2002).  The thick contour is the 3$\sigma$
confidence level for three degrees of freedom, at a $\Delta\chi^2$ of
14.2 above the minimum. The intrinsic flattening $q$ of the models is
indicated in the lower-right corner of each panel. Left: model fits to
the Joseph et al.\ (2001) {\tt STIS} major-axis data and the {\tt
SAURON} integral-field measurements over $9''\times 11''$ sampled with
0\farcs28 pixels in 0\farcs95 seeing.  This allows accurate
measurement of $M/L$, $M_{\rm BH}$ and $i$.  Right, model fits which
include again the {\tt STIS} major axis-data, but now only the {\tt
SAURON} measurements along 4 slits (major and minor axis, and at $\pm
45^\circ$).  This shows that the traditional kinematic coverage
provides almost no constraint on $i$, and that the resulting
uncertainties on the inferred values of $M/L$ and $M_{\rm BH}$ are
significantly larger.\looseness=-2 }
\vskip -8.0truept
\end{figure}

\subsection{The E3 galaxy M32} 
 
M32 is a high-surface brightness inactive E3 companion of M31, and has
long been suspected of harboring a central black hole of about
$3\times 10^6 M_\odot$ (e.g., Tonry 1984). This dominates the observed
kinematics inside 0\farcs15, and as a result a reliable mass
measurement had to await {\tt FOS} and the development of general
machinery for the construction of axisymmetric dynamical models with
the full range of orbital anisotropies (van der Marel et al.\
1998). Verolme et al.\ (2002) compared the {\tt STIS} major axis
kinematics (Joseph et al.\ 2001), as well as {\tt SAURON}
integral-field spectroscopy of the inner $9''\times 11''$ of M32
(Bacon et al.\ 2001b; de Zeeuw et al.\ 2002) with general axisymmetric
dynamical models, varying stellar mass-to-light ratio $M/L$, black
hole mass $M_{\rm BH}$ and the inclination $i$. All three parameters
are well-constrained, with $M_{\rm BH}=(2.5\pm0.5) \times 10^6
M_\odot$ (3$\sigma$-error), $M/L=1.85\pm0.15 \, M_\odot/L_\odot$ in
the $I$-band, and $i=70^\circ \pm 5^\circ$ (Figure 2, left panel).
Experiments showed that the traditional approach of using ground-based
kinematics along a few slits significantly degrades the measurement
accuracy: it does not allow meaningful determination of $i$ and as a
result causes a larger uncertainty on $M_{\rm BH}$ (Figure 2, right
panel). Even in the case of an object as dynamically simple as M32,
two-dimensional coverage is of key importance.

\subsection{Kinematics of gas or stars?} 
 
Much work has been done on obtaining kinematics of emission-line gas
of nearby nuclei (Kormendy \& Gebhardt 2002).  This is attractive as
it allows kinematic measurements with only a few HST orbits at
excellent spatial resolution. Often three parallel slits are used to
mimic integral-field capability. The results are not always easy to
interpret, as the gas is rarely in a simple disk, and even if it is,
the kinematics may be asymmetric, or the gas velocity dispersion so
large that modeling becomes problematic (e.g., Ho et al.\ 2002).

Black hole masses and nuclear properties derived independently from
the kinematics of gas and stars are available for only a few nearby
galaxies. Cappellari et al.\ (2002) carried out a detailed comparison
for the E3 galaxy IC 1459. Dynamical modeling of the stellar
kinematics (using {\tt STIS} and multiple-position angle ground-based
data) allows a decomposition/identification of the counter-rotating
stellar core in phase space, and shows that it is a disk with a mass
of about $3 \times 10^9 M_\odot$. The best-fit model has a black hole
mass of about $2.6\times 10^9 M_\odot$, somewhat larger than the value
expected from the $(M_{\rm BH}, \sigma)$-relation. Models of the gas
velocities suggest $M_{\rm BH} \approx$$3.5\times 10^8 M_\odot$ if the
gas is in circular motion in a principal plane, while simple models of
the velocity dispersion suggest $M_{\rm BH} \approx$$1\times 10^9
M_\odot$ for a spherical distribution of the gas.  This leaves
significant uncertainties about the reliability of the gas as a tracer
of the galaxy gravitational potential. Studies of NGC4335 (Verdoes
Kleijn et al.\ 2002), and NGC4697 (Gebhardt priv.\ comm.) reach
similar conclusions. While simple modeling of the observed gas
kinematics can probably be trusted when the gas has a smooth and
regular morphology, symmetric kinematics, and low velocity dispersion
(e.g., Barth et al.\ 2001), in many nuclei this will not be the
case. 


Much effort has gone into obtaining stellar kinematics with {\tt STIS}
for nearby galactic nuclei (e.g., Pinkney et al.\ 2002). These studies
have typically been restricted to objects with velocity dispersions
larger than about 140 km/s, and cusped luminosity profiles. In some
cases the 0\farcs2 slit is required to reach sufficient
signal-to-noise. In combination with multi-position-angle or
integral-field ground-based stellar kinematics, these measurements
provide reliable black hole masses (e.g., Gebhardt et al.\ 2002;
Verolme et al.\ 2002). {\tt STIS} cannot probe the nuclei of the giant
ellipticals in Virgo because of their modest surface brightness, and
cannot resolve the black hole sphere of influence in low-luminosity
ellipticals, because at 15 Mpc its radius is smaller than 0\farcs1.

\subsection{Active galaxies} 

Most nearby galaxies contain a quiescent or only weakly active nucleus
(Ho et al.\ 1997). There are few powerful nuclei within 20 Mpc: Cen A
at $\approx$3 Mpc (but its nucleus is heavily obscured), as well as
classical Seyferts such as NGC1068, and FR~I radio galaxies such as
M84 and M87 in the Virgo cluster. More powerful active nuclei appear
at even larger distances, with Cyg A, the classical FR~II at $\sim$200
Mpc. 3C273, the nearest quasar, is another factor 2.5 or so more
distant.\looseness=-2

HST imaging by Verdoes Kleijn et al.\ (1999) of a complete sample of
21 northern nearby FR~I galaxies (18 of which are beyond 40 Mpc)
revealed that all of them have nuclear emission-line gas, as well as
dust disks or dust lanes. Comparison with imaging of inactive galaxies
shows that the trigger of activity must lie on scales smaller than
resolved by {\tt WFPC2}, even in the nearest active nuclei.  {\tt
STIS} emission-line kinematics is now available for these FR~I objects
but, as we have seen above, the derived $M_{\rm BH}$ will have to be
treated with caution, in particular because the gas kinematics may be
infuenced by non-gravitational motions (in- and outflows). Beyond
40~Mpc the spatial resolution is insufficient to detect all but the
largest black holes. Measurement of the stellar kinematics in nearby
active nuclei is difficult, as these often reside in low-surface
brightness giant ellipticals, and contain a bright nuclear
emission-line spike.  Increased resolution in both imaging and
spectroscopy is required.

\section{Requirements for next steps} 

HST Studies of nearby galactic nuclei have raised interesting
questions, including: why are nuclear and global properties
correlated, what is the role of nuclear disks and asymmetries, do
galaxies evolve dynamically from the inside out on interesting time
scales, why is only a small fraction of the black holes currently
active, how does the $M_{\rm BH}$ mass function evolve with time?  In
order to answer these, we need to extend the studies that are
currently possible only for Local Group galaxies to a representative
sample. This leads to four requirements:

\begin{itemize}
\itemsep -4truept
\item Integral-field spectroscopy to obtain the two-dimensional
      kinematics (and line-strengths), which provides key constraints
      not only for asymmetric nuclei such as the one in M31, but also
      in `simple' galaxies such as M32;
\item Sufficient sensitivity to be able to measure the stellar 
      absorption-line kinematics over the full range of central surface 
      brightness seen in Virgo galaxies, in order to have a reliable tracer 
      of the gravitational potential; 
\item Increased spatial resolution, to probe the nuclei of the smaller
      galaxies in Virgo, and to carry out a representative census to
      larger distances which also contains the rarer types, in
      particular the radio-loud galaxies.
\item A coronographic capability to probe near bright central spikes.
\end{itemize}

\section{Prospects on the ground} 
 
Galactic nuclei are interesting targets for AO-assisted studies from
the ground. The {\tt OASIS} results on M31 in the $I$-band shown in
Figure 1 provide a case in point. This instrument is being upgraded,
and will move to the WHT on La Palma in 2003. The enhanced throughput,
and a laser guide star (planned for 2005) should provide the stellar
kinematics in objects like M87 with unprecedented spatial resolution.

Integral-field spectrographs operating at near-infrared wavelengths
will soon be available on 8m class telescopes. Examples include {\tt
SINFONI} on the VLT and {\tt NIFS} on Gemini, with first light for
both expected in 2004. These instruments allow measurement of the
stellar kinematics through observations of the CO bandhead at
2.3$\mu$m. With the tenfold increase in collecting area over HST, and
the potential to provide spectra at a spatial resolution of better
than 0\farcs1 at 2$\mu$m, this will allow a significant step beyond
{\tt STIS} in the study of normal and dusty objects, including some of
the nearest active nuclei.

Adaptive optics in the visible range on 8m telescopes can in principle
improve on the current HST resolution by a factor two or more over a
small field of view. This will be hard to achieve in practice, but
there is much ongoing effort in this direction, and galactic nuclei
are among the natural targets. This will not only involve imaging but
also spectroscopy. An example is provided by the proposed second
generation VLT instrument {\tt MUSE} (PI Bacon). While its main aim
will be MC(AO) assisted panoramic integral-field spectroscopy of
ultra-deep fields in the $I$ and $R$ bands, {\tt MUSE} will have a
mode for spectroscopy at the highest spatial resolution provided by
adaptive optics in a small field. Phase A is in progress, and the
instrument could be on the VLT by 2007.
 
Development of interferometry at infrared wavelengths is being pursued
actively at Keck and at ESO (VLTI). Interferometry is also an
important component of the planned scientific use of the LBT.
Galactic nuclei (including the Galactic Center) are among the main
drivers to go faint and reach $K$ of 20 mag. The anticipated VLTI
spatial resolution of order 0\farcs015 at 10 $\mu$m would allow
imaging of the dusty broad-line regions of the nearest active nuclei
at these wavelengths. Obtaining spectroscopic information on these
scales would be very exciting, but is an extremely challenging goal.
 
Design studies are underway for optical telescopes with filled
apertures ranging from 30 (CELT and GSMT) to 100m (OWL). If (MC)AO can
be made to work at near-infrared and eventually at optical
wavelengths, then these extremely large telescopes will provide the
increase in resolution over HST, combined with sufficient sensitivity,
to carry out studies of galactic nuclei satisfying the requirements
outlined in section 3.  A 30m telescope would allow detailed study of
nuclei in most Hubble types. Larger apertures (70--100m) operating at
visible wavelengths would in principle allow detailed studies out to
beyond Virgo with the sensitivity and resolution now achievable only
for Local Group galaxies. Furthermore, because the angular diameter
versus distance relation has a minimum at $z=1$, these ELT's could
make it possible to estimate black hole masses in large galaxies
throughout the Universe from emission-line kinematics.

\vfill\eject

\section{Prospects in space} 

The first question to consider is whether significant progress could
already be made by equipping HST with an integral-field spectrograph
for near-UV and optical wavelengths.  Efficient high-throughput
designs exist, but this option would require an additional servicing
mission. The reward would be the enhancement of the spectroscopic
efficiency of HST by about a factor of 50 beyond {\tt STIS}, including
a relaxation of the roll-angle requirements on the observatory.
Applications would range from resolved spectroscopy of Io and of
proplyds in star-forming regions to resolved kinematics of
gravitationally lensed arcs at high redshift. For galactic nuclei,
this would provide exquisite capabilities for studying the nuclei of
Local Group galaxies, and for measuring the gas kinematics in many
nuclei.  It would also enlarge the sample for which stellar kinematics
can be measured with sufficient resolution.\looseness=-2

The current plans have NGST as a 6m class observatory, with
spectroscopic capabilities in the near and mid-infrared. An
integral-field capability on {\tt NIRSPEC} would allow study of normal
and dusty nearby nuclei, provided it has a spectral resolution of a
few thousand. The main gain over HST would not be spatial resolution
(because of the longer wavelengths) but an order of magnitude in
sensitivity, so that one might obtain stellar kinematics of giant
ellipticals at 0\farcs1 resolution. {\tt MIRI} integral-field
spectroscopy will provide emission-line diagnostics of dusty
starbursts and active nuclei, and may probe the nuclear stellar
populations of high-$z$ galaxies as the CO bandhead moves to the
mid-IR for $z > 1.5$.\looseness=-2
 
A 6-8m class space telescope with integral-field capabilities in the
optical (and perhaps near-UV) would allow stellar absorption-line
spectroscopy of many galactic nuclei at 0\farcs1 resolution or better,
and a sufficient gain over HST in sensitivity that the giant
ellipticals in Virgo can be studied at this resolution. A coronograph
would allow unique studies of active nuclei (e.g., Seyferts).  This
would be a truly fantastic resource, which would benefit broad areas
of astronomy. However, {\em for galactic nuclei} it might not provide
a huge gain over the results expected from HST imaging combined with
AO-assisted ground-based integral-field spectroscopy on 8m class
telescopes, except in UV diagnostics.  A space telescope of
significantly larger aperture would provide a major leap in this area,
and allow absorption-line studies in a representative sample of galaxy
types with sufficient spatial resolution and sensitivity, and
emission-line studies to the highest redshifts.

\section{Conclusions} 
 
The combination of HST imaging and ground-based integral-field
spectroscopy is very powerful for the study of the nuclei of normal
and active galaxies, and provides a compelling argument for
integral-field capabilities in space.  Much progress in our
understanding of galactic nuclei is expected on the ground in the next
decade, and this may continue {\em if} near-diffraction limited
observations become possible on telescopes with apertures of 30m or
larger. While the study of galactic nuclei is not the main science
driver for a 6-8m class space telescope, even a smaller aperture
equipped with an integral-field spectrograph would yield a rich
scientific harvest.

\acknowledgments It is a pleasure to thank Eric Emsellem for providing
Figure 1, and to thank him, Roland Bacon, Wilfried Boland, Michele
Cappellari, Gijs Verdoes Kleijn and Ellen Verolme for comments on the
manuscript. ESA support for participation in this conference is
gratefully acknowledged.

\end{document}